# A Protection-Interoperable Fault Ride-Through Control for Grid-Forming Inverters

Yifei Li, Student Member, IEEE, Heng Wu, Senior Member, IEEE, Xiongfei Wang, Fellow, IEEE

*Abstract*—Differing from synchronous generators (SGs), grid-forming inverter-based resources (GFM-IBRs) exhibit rapid variations in their output impedances during transmission line faults due to the overcurrent limitation. As a result, the source dynamics during the fault period deviate significantly from those under pre-fault conditions. This fundamental difference alters the fault responses of incremental quantities, thereby jeopardizing the reliability of the supervising elements in protective relays that are based on these quantities. To address this challenge, a protection-interoperable fault ride-through (FRT) method for GFM-IBRs is proposed. This method dynamically adjusts power control of GFM-IBRs in response to the changes in output impedance, effectively mitigating variations in source dynamics and thereby preserving the reliability of incremental quantity-based supervising elements. This method also ensures effective overcurrent limitation and transient stability of GFM-IBRs. Controller hardware-in-the-loop (CHIL) and experimental tests validate the effectiveness of the proposed method.

## I. Introduction

GRID-forming inverter-based resources (GFM-IBRs) are treated as a promising substitute technology for synchronous generators (SGs) in future power systems with high penetration of renewable energy power generation [1], [2]. However, deploying GFM-IBRs at large scale can significantly alter fault characteristics of power systems, which may further affect the reliability of the protective relays that are designed for legacy SG-based power systems [3], [4].

Supervising elements, including directional and phase selection elements, are standard elements that are integrated into protective relays to ensure tripping occurs only for faulted phases during a forward fault [3], [5], thereby enhancing the resilience and reliability of power systems [6], [7]. To eliminate the impact of load conditions, the phase angle characteristics of incremental quantities, i.e., the incremental voltage and current, are commonly employed in supervising elements for assessing the faulted phases and the direction of the fault [8], [9], [10]. The operation of these incremental quantity-based supervising elements is built on the compensation theory, which assumes that the variations of source dynamics, including the internal voltage source (IVS) and output impedance dynamics, are negligible during the fault period when compared to pre-fault conditions [11], [12]. For GFM-IBRs, the variations in IVS dynamics are negligible under the typical slow-timescale power control [2]. However, due to overcurrent limitations, the GFM-IBRs undergo rapid changes in their output impedances during the fault period, which is different from SGs [3]. This distinction results in the source dynamics during the fault period differing significantly from those before the fault. Consequently, the phase angle characteristics of incremental quantities are altered, jeopardizing the reliability of the incremental quantity-based supervising elements.

To ensure reliable operations of incremental quantity-based supervising elements, either the protective algorithm or the control scheme of the IBR should be modified [13]. Given the continuous evolution of IBR characteristics driven by updates of grid codes, one promising solution for the modification of protective algorithm is to make it decoupled from the source characteristics [13], [14]. For instance, the SEL-T400L employs traveling waves to determine the direction of the fault and time-domain quantities to identify the faulted phases [15]. While these methods operate mostly independently of the source characteristics, they require a high sampling rate of 1 MHz, posing challenges for hardware implementation [15]. Further, updating all these protective elements increases the economic cost [6], [16]. Therefore, implementing protection-interoperable control to ensure the compatibility of GFM-IBRs with the conventional supervising elements presents a cost-effective and attractive solution [17].

In [7], the current reference generation of IBRs is adjusted according to the tripping criteria of the directional elements to enhance reliability. Similarly, in [6] and [18], the control schemes are adjusted according to the tripping criteria of the phase selection elements. However, modifying control schemes to be adapted to one protective element may adversely impact the reliability of other elements. The unified requirements for both directional and phase selection elements are thus needed to effectively modify the control schemes of IBRs. Moreover, most previous studies focus on modifying the grid-following (GFL) control [4]. Unlike GFL-IBRs, GFM-IBRs operate as IVSs behind output impedances, exhibiting high stability in weak grids [2], [19]. However, the protection-interoperable control methods for GFM-IBRs remain largely underexplored [4].

Beyond meeting the requirements of protective relays, IBRs should also reliably ride through grid faults while accounting for overcurrent limitations and transient stability [20], [21]. In [1], the steady-state fault current is limited by adjusting the power references, and the transient fault current is limited by using the virtual resistor control. The effects of various current limiting controls (CLCs) on transient stability and dynamic behavior of IBRs are summarized in [21]. An adaptive fast/slow control method is introduced in [2], which guarantees the transient stability of GFM-IBRs. While the CLC and the transient stability are thoroughly studied in [1], [2], [20], and

[21], these works do not consider the impact of control strategies on the reliability of supervising elements.

To bridge the above-mentioned gaps, this paper proposes a protection-interoperable FRT method for GFM-IBRs, which assures reliable operations of the incremental quantity-based supervising elements and effectively meets the requirements of overcurrent limitations and transient stability of GFM-IBRs, all are essential control targets. **In contrast, the transient stability is often overlooked in prior studies that primarily focus on the reliable operation of protective relays [6], [7], [18], while the protective relays are seldom mentioned in the studies on transient stability** [2]. The key aspects of the protection-interoperable FRT method are as follows:

1) The power control of IBRs is modified such that source dynamics, including the dynamics of IVS and output impedance, in the fault sequence network remain equivalent to those in the pre-fault operation, thereby maintaining reliable operations of supervising elements that are based on incremental quantities.
2) The modified power control operates in an open-loop manner and is activated exclusively during the operation of the supervising elements, whereas at all other times, the closed-loop power control is employed to maintain synchronization of the IBR with the grid. The transition between these two modes is governed by the switching criteria, which are designed based on the comparison of current magnitude against predefined thresholds and the operating duration of supervising elements.
3) The requirements for preventing overcurrent during IBR operation under the modified power control, imposed on the output impedance shaped by the CLC, are established.

The advantages of the proposed method are listed below:
1) The method ensures reliable operations of both incremental quantity-based directional and phase selection elements.
2) With the open-loop power control adopted initially after the fault inception, and the adaptive fast/slow control [2] used during fault recovery, the proposed method assures the transient stability of GFM-IBRs.

Lastly, the controller hardware-in-the-loop (CHIL) tests and experiments validate the effectiveness of the proposed method.

## II. PRECONDITIONS OF SUPERVISING ELEMENTS

This section briefly reviews the operating principles of supervising elements used in power system protection, based on which the preconditions for their reliable operation are summarized.

### A. Principle of Supervising Elements

Fig. 1 shows the single-SG infinite bus system. Here, $v$, $i$, and $v_e$, $i_g$ represent the voltages and currents at the bus 1 and the bus 2, respectively. $Z_l$ denotes the transmission line impedance. Fault points in the reverse and forward directions are marked as $F_x$ and $F_y$, respectively. $R_1$ and $R_2$ are the distance relays that incorporate the distance elements and supervising elements, including directional and phase selection elements. The *bcg* (phase *b* to phase *c* to ground) fault at $F_y$ is used as an example to illustrate the principle of supervising elements.

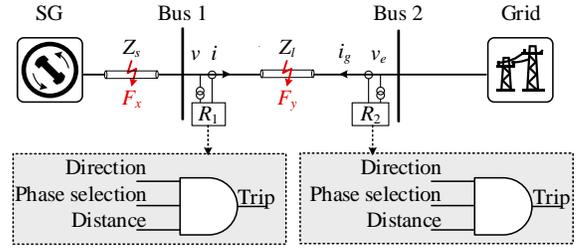

Fig. 1 Single-SG infinite bus system.

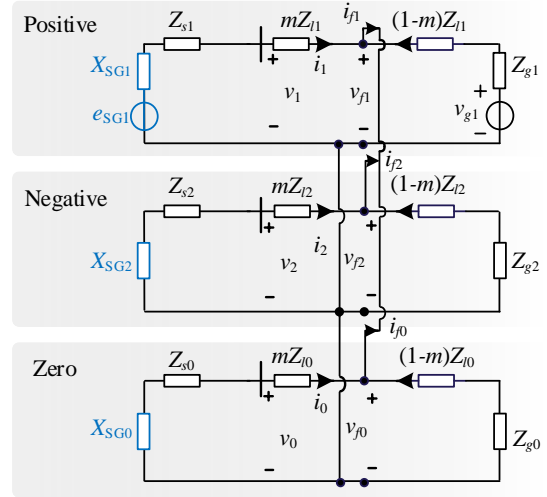

Fig. 2 Sequence network under a *bcg* fault in the single SG-based power system.

### 1) Principle of directional element

Fig. 2 depicts the corresponding fault sequence network of Fig. 1, where $v_{SG1}$ denotes the internal voltage of the SG. $X_{SG}$ and $Z_g$ represent the output impedances of the SG and grid, respectively. The symbol $m$ denotes the fault location, and the subscript '*f*' denotes the quantities at the fault location. The subscripts 0, 1, and 2 represent zero-, positive-, and negative-sequence quantities, respectively. Applying Kirchhoff's law to the sequence network in Fig. 2, the angle differences between voltages and currents of negative-sequence quantities ($\varphi_2$) and zero-sequence quantities ($\varphi_0$) are expressed as

$$\begin{cases} \varphi_2 = \angle \frac{v_2}{i_2} = \angle - \underbrace{(Z_{s2} + jX_{SG2})}_{Z_{e2}} \\ \varphi_0 = \angle \frac{v_0}{i_0} = \angle - \underbrace{(Z_{s0} + jX_{SG0})}_{Z_{e0}} \end{cases} \quad (1)$$

where $Z_{e2}$ and $Z_{e0}$ represent the effective impedances. Due to the fact that the output impedances of the SG and transmission line are highly inductive [22], it is known from (1) that a forward fault, e.g., the fault occurs at $F_y$, results in $\varphi_2 \approx \varphi_0 \approx -90°$. In contrast, in the case of a reverse fault, e.g., the fault occurs at $F_x$, the fault current holds an opposite phase angle compared to that during the forward fault, and the effective impedances are formed by the grid-side impedances [23], yielding $\varphi_2 \approx \varphi_0 \approx 90°$. Therefore, the values of $\varphi_2$ and $\varphi_0$ can be used to identify the direction of the fault.

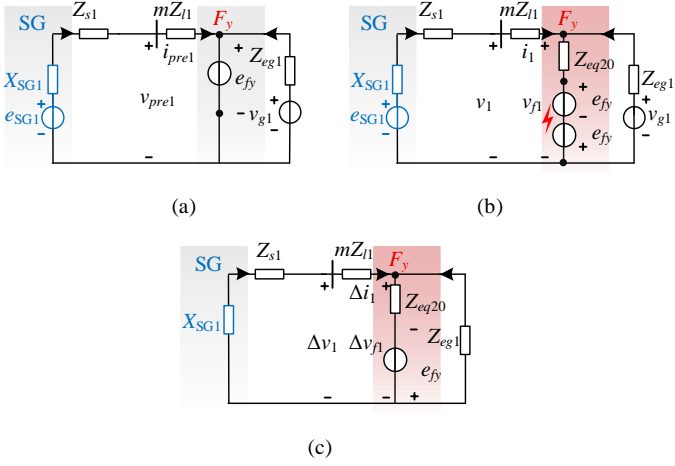

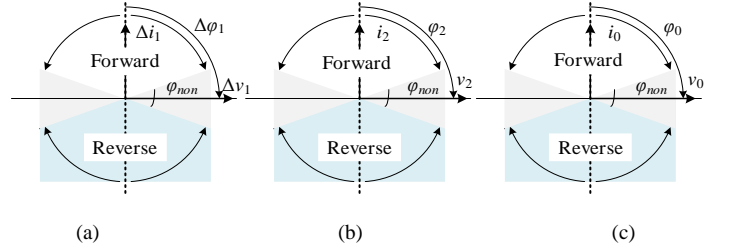

Fig. 4 Phase angle characteristics of directional elements. (a) $\Delta\varphi_1$. (b) $\varphi_2$. (c) $\varphi_0$.

Following (5), $\Delta v_1$ and $\Delta i_1$ are obtained through the subtraction between the fault quantities ($v_1$, $i_1$) and pre-fault quantities ($v_{pre1}$, $i_{pre1}$), and the angle difference between $\Delta v_1$ and $\Delta i_1$ is given by

$$\Delta\varphi_1 = \angle - \frac{\Delta v_1}{\Delta i_1} \approx \angle - \underbrace{(jX_{\text{SG1}} + Z_{s1})}_{\Delta Z_{e1}} \qquad (6)$$

where $\Delta Z_{e1}$ is the effective impedance for the incremental quantities. During a forward fault, based on (6), $\Delta\varphi_1 \approx -90°$. During a reverse fault, $\Delta i_1$ holds an opposite phase angle to the forward fault, resulting in $\Delta\varphi_1 \approx 90°$. Therefore, the value of $\Delta\varphi_1$ can also be used to identify the direction of the fault.

Furthermore, in Fig. 3 (c), the $e_{\text{SG1}}$ and $v_{g1}$ are removed. Therefore, by using $\Delta v_1$ and $\Delta i_1$ in Fig. 3 (c), the adverse impact of load conditions on the performance of the supervising elements is eliminated.

Fig. 4 illustrates the principles of the directional elements. The angle differences between voltage and current for the incremental quantities ($\Delta\varphi_1$) in Fig. 4 (a), the negative-sequence quantities ($\varphi_2$) in Fig. 4 (b), and the zero-sequence quantities ($\varphi_0$) in Fig. 4 (c) are used to identify the direction of the fault. $\Delta\varphi_1$, $\varphi_2$, and $\varphi_0$ fall within the reverse or forward zones based on the corresponding direction of the fault. Additionally, the non-tripping zones ($\varphi_{non}$) are defined within a range of 30° to 60° to improve the reliability of the directional elements [24].

### 2) Principle of phase selection elements

A one-to-one mapping exists between the angle differences at the fault location [$\delta_{f21}=\angle i_{f2}-\angle i_{f1}$, $\delta_{f20}=\angle i_{f2}-\angle i_{f0}$] and the fault types, forming the foundation for faulted phases identification [10]. However, the relay cannot directly measure the quantities at the fault location. When the output impedances of the sources are highly inductive, the phase angles of the sequence quantities at the relay-assembled point correspond to those at the fault location for negative- and zero-sequence quantities, i.e., $\angle i_2 = \angle i_{f2}$, and $\angle i_0 = \angle i_{f0}$, enabling identification of faulted phases. For further details, refer to [10] and [11].

However, due to the impact of load conditions, $\angle i_1$ does not necessarily match $\angle i_{f1}$. Since the incremental quantity remains unaffected by load conditions, i.e., $\angle \Delta i_1 = \angle i_{f1}$, it can be used to identify the faulted phases [10].

Fig. 5 shows the principles of the phase selection elements. The angle difference between $i_2$ and $\Delta i_1$ ($\Delta\delta_{21}$) shown in Fig. 5 (a), and the angle difference between $i_2$ and $i_0$ ($\delta_{20}$) shown in Fig. 5 (b) are used simultaneously to identify faulted phases. To enhance reliability, the fault type zones for $\Delta\delta_{21}$ and $\delta_{20}$ are

Fig. 3 Illustration of the compensation theory in the SG-based system. (a) Pre-fault circuit. (b) Fault sequence network. (c) Pure-fault sequence network.

However, in the event of a symmetrical fault, the negative- and zero-sequence quantities are absent, leaving only the positive-sequence quantities available for the supervising elements. Nevertheless, the positive-sequence quantities are affected by load conditions [10]. Consequently, the incremental quantities derived based on compensation theory are used instead.

Fig. 3 illustrates the compensation theory in the SG-based system, where $Z_{eg1}=(1-m)Z_{l1}+Z_{g1}$. $Z_{eq20}$ is the equivalent parallel impedance for negative- and zero-sequence circuits. $e_{fy}$ and $v_{f1}$ are the pre-fault and fault voltages at the point $F_y$, respectively. The subscript '*pre*' in Fig. 3 (a) stands for the pre-fault quantities. For analysis of protective relays, the IVS ($e_{\text{SG1}}$) and output impedance ($Z_{\text{SG1}}$) of SGs in the fault sequence network are generally assumed to be similar to those in the pre-fault circuit [3], [11], [22], as shown in Fig. 3 (a) and (b).

Applying nodal analysis to the pre-fault circuit [Fig. 3 (a)] and fault sequence network [Fig. 3 (b)], the voltages at the fault location are separately derived by

$$e_{fy} = \frac{e_{\text{SG1}}(Y_{\text{SG1}} + Y_{s1} + mY_{l1}) + v_{g1}Y_{eg1}}{Y_{\text{SG1}} + Y_{s1} + mY_{l1} + Y_{eg1}} \qquad (2)$$

$$v_{f1} = \frac{e_{\text{SG1}}(Y_{\text{SG1}} + Y_{s1} + mY_{l1}) + v_{g1}Y_{eg1}}{Y_{\text{SG1}} + Y_{s1} + mY_{l1} + Y_{eg1} + Y_{eq20}} \qquad (3)$$

where the symbol $Y$ stands for admittance. Subtracting (2) from (3), yields

$$\Delta v_{f1} = v_{f1} - e_{fy}$$
$$= \frac{-e_{fy}Y_{eq20}}{Y_{\text{SG1}} + Y_{s1} + mY_{l1} + Y_{eg1} + Y_{eq20}} \qquad (4)$$

Fig. 3 (c) presents the pure-fault sequence network derived based on (4). Applying Kirchhoff's law to the circuit in Fig. 3 (c), the $\Delta i_1$ and $\Delta v_1$ are expressed as

$$\begin{cases} \Delta i_1 = -\frac{\Delta v_{f1}}{jX_{\text{SG1}} + Z_{s1} + mZ_{l1}} \\ \quad = -\underbrace{\frac{v_{f1} - e_{\text{SG1}}}{jX_{\text{SG1}} + Z_{s1} + mZ_{l1}}}_{i_1} - \underbrace{\left(-\frac{e_{fy} - e_{\text{SG1}}}{jX_{\text{SG1}} + Z_{s1} + mY_{l1}}\right)}_{i_{pre1}} \\ \Delta v_1 = -\Delta i_1(jX_{\text{SG1}} + Z_{s1}) \\ \quad = \underbrace{e_{\text{SG1}} - i_1(jX_{\text{SG1}} + Z_{s1})}_{v_1} - \underbrace{(e_{\text{SG1}} - i_{pre1}(jX_{\text{SG1}} + Z_{s1}))}_{v_{pre1}} \end{cases} \qquad (5)$$

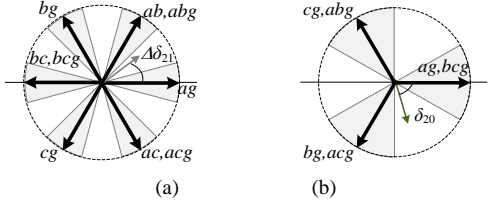

Fig. 5 The principle for PSEs. (a) $\Delta\delta_{21}$. (b) $\delta_{20}$.

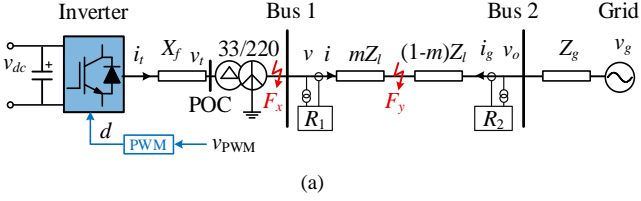

(a)

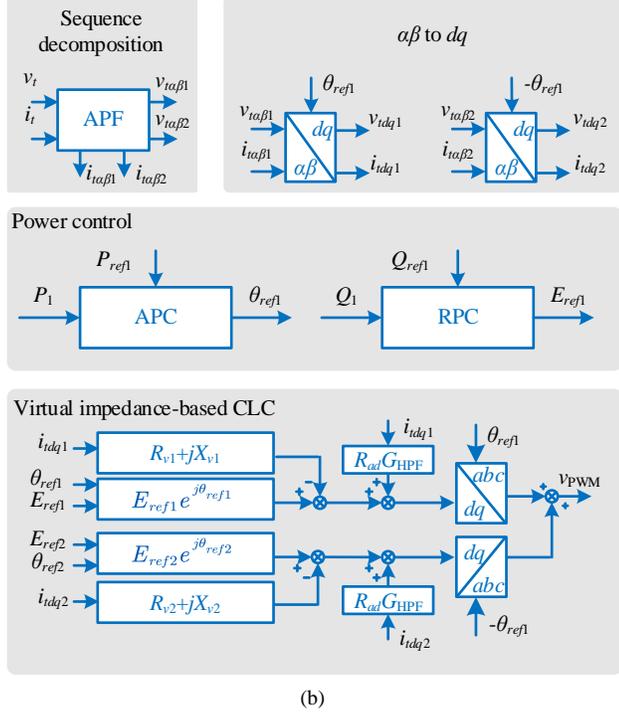

(b)

Fig. 6 Main circuit and control schemes of the single-IBR infinite bus system. (a) Main circuit. (b) Control loops.

extended by ±15° and ± 30°, respectively [25].

### B. Preconditions

Based on the above analysis, the preconditions for the supervising elements are listed below:
a) The effective impedance is highly inductive.
b) The difference in source dynamics, including IVS and output impedance dynamics, between the fault sequence network and the pre-fault circuit is negligible.

## III. IMPACT OF GFM-IBRs ON SUPERVISING ELEMENTS

### A. System Description

Fig. 6 (a) shows the main circuit of the single-IBR infinite bus system. $v_t$ and $i_t$ are the voltage and current at the point of coupling (POC), respectively. $X_f$ represents the filter. A constant DC-link voltage ($v_{dc}$) is assumed, as it is typically maintained by the front-end converter or an energy storage unit [1].

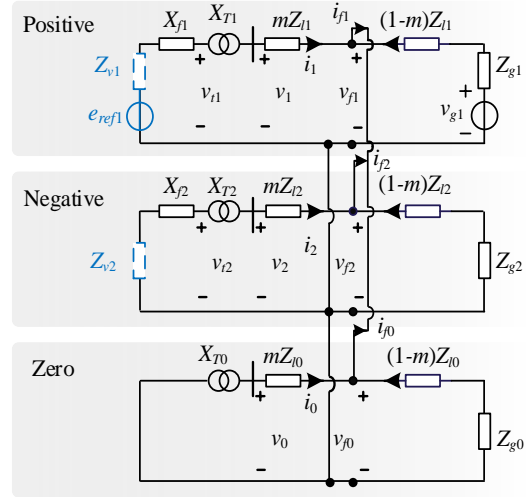

Fig. 7 Sequence network under a $bcg$ fault in GFM-IBR-based systems.

Fig. 6 (b) illustrates the control scheme for the GFM-IBR, where the all-pass filter (APF) is utilized to decompose the positive- and negative-sequence quantities, considering the asymmetrical faults at the transmission line [26]. The power control consists of active power control (APC) and reactive power control (RPC). The APC is used to generate the phase angle reference $\theta_{ref1}$ for $dq$ transformation, while the RPC is utilized to generate the positive-sequence voltage magnitude reference ($E_{ref1}$). To ensure a balanced voltage, the negative-sequence voltage magnitude reference is set as zero, i.e., $E_{ref2}=0$ [27].

To prevent the failure of power semiconductor devices from overcurrent, the virtual impedance-based CLC method is implemented. In this method, the active damping resistor $R_{ad}$ is used to dampen the synchronous resonance introduced by the power control. Moreover, a high-pass filter ($G_{HPF}$) is cascaded with $R_{ad}$ to eliminate its impact on the steady-state power control performance [2].

### B. Impact Analysis

#### 1) X/R ratio of the virtual impedance

Fig. 7 presents the sequence network of Fig. 6 (a) during a $bcg$ forward fault, where $e_{ref1}=E_{ref1}\angle\theta_{ref1}$, and $X_T$ is the leakage reactance of the transformer. Applying Kirchhoff's law to the circuits in Fig. 7, the angle differences between voltages and currents of negative-sequence quantities ($\varphi_2$) and zero-sequence quantities ($\varphi_0$) are expressed as

$$\begin{cases} \varphi_2 = \angle\dfrac{v_2}{i_2} = \angle - \underbrace{(n^2 Z_{v2} + jn^2 X_{f2} + jX_{T2})}_{Z_{e2}} \\ \varphi_0 = \angle\dfrac{v_0}{i_0} = \angle - \underbrace{(jn^2 X_{f0} + jX_{T0})}_{Z_{e0}} \end{cases} \quad (7)$$

where $n$ is the turns ratio of the transformer. Based on (7), $Z_{e2}$ is affected by the virtual impedances. Therefore, it may not be highly inductive, which may cause malfunctions of supervising elements depicted in Fig. 4 and Fig. 5.

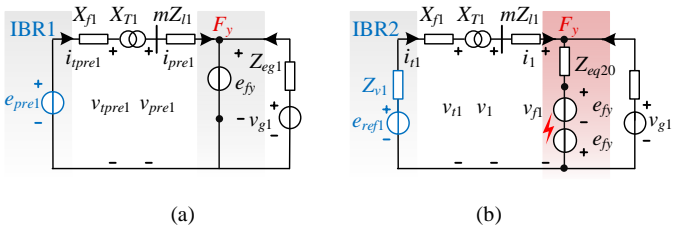

(a)          (b)

Fig. 8 Illustration of the source dynamic in IBR-based systems. (a) Pre-fault circuit. (b) Fault sequence network.

To address this issue, a highly inductive virtual impedance-based CLC method is implemented to ensure $Z_{v2}$ remains highly inductive. Consequently, $Z_{e2}$ is maintained as highly inductive. However, the performance of incremental quantity-based supervising elements is affected by the source dynamics and should be further investigated.

### 2) Source dynamics

Fig. 8 illustrates the source dynamics, including IVS and output impedance dynamics, where IBR1 and IBR2 denote the IBR models before and during the fault, respectively. Here, $e_{pre1}$ represents the IVS of the IBR before the fault. Fig. 8 (a) shows that the overcurrent limitation is not triggered before the fault, and the output impedance of the IBR1 is 0. During the fault, with the typically slow-timescale power control, the IVS remains approximately equal to its pre-fault value, i.e., $e_{ref1} \approx e_{pre1}$ [2]. However, when the virtual impedance-based CLC is triggered, an additional impedance $Z_{v1}$ is introduced as the output impedance of the IBR2, as depicted in Fig. 8 (b). Therefore, unlike the SG, the output impedance of the IBR undergoes rapid changes, resulting in mismatched source dynamics between IBR2 in Fig. 8 (b) and IBR1 in Fig. 8 (a). Consequently, the assumption underlying the compensation theory—that the source dynamics remain nearly unchanged—no longer holds.

Directly subtracting the pre-fault quantities in the circuit of Fig. 8 (a) from the fault quantities in the sequence network of Fig. 8 (b), the effective impedance for the incremental quantities is expressed as

$$\Delta Z_{e1} = -\frac{\Delta v_1}{\Delta i_1} = -\frac{v_1 - v_{pre1}}{i_1 - i_{pre1}}$$
$$= (jn^2 X_{f1} + jX_{T1}) + \underbrace{\frac{ne_{pre1} - ne_{ref1} + n^2 i_1 Z_{v1}}{i_1 - i_{pre1}}}_{Z_{ad}} \quad (8)$$

Thanks to the slow-timescale power control, $e_{ref1} \approx e_{pre1}$. However, the phase angle of $i_1$ does not necessarily match that of $i_1$-$i_{pre1}$. Hence, $Z_{ad}$ cannot be ensured highly inductive. Consequently, based on (8), the expected highly inductive $\Delta Z_{e1}$ cannot be directly determined by subtracting pre-fault quantities from the fault quantities, which can adversely impact the performance of the incremental quantity-based supervising elements.

### C. Challenge

The challenge posed by GFM-IBRs to the performance of supervising elements is, therefore, summarized below:
a) Unlike SGs, GFM-IBRs exhibit significantly different source dynamics during fault conditions compared to pre-fault conditions. Consequently, the incremental quantity-based supervising elements cannot be directly applied.

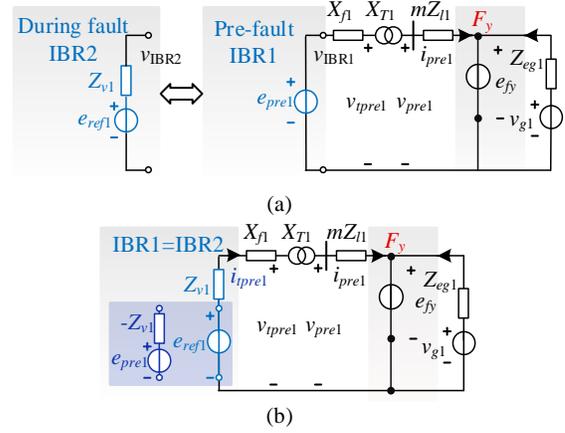

(a)

(b)

Fig. 9 Illustration of the equivalent dynamics in the pre-fault circuit. (a) Replacement of the IBRs. (b) Equivalent pre-fault circuit.

## IV. PROTECTION-INTEROPERABLE FRT CONTROL

This section presents first the general idea of protection-interoperable FRT control for GFM inverters, developed to address the identified challenge. Then, the pure-fault circuit for the incremental quantities under this control is derived to demonstrate its effectiveness.

### A. General Idea

In SG-based systems, the IVS and output impedance dynamics in the fault sequence network remain nearly unchanged from those in the pre-fault circuit, allowing the expected incremental quantities to be derived by subtracting the pre-fault quantities from the fault quantities. As a result, the incremental quantity-based supervising elements are applicable.

Inspired by the equivalence between SG dynamics in fault sequence networks and pre-fault circuits, a similar principle can be applied to IBR-based systems. Specifically, if IBR2 exhibits equivalent IVS and output impedance dynamics as IBR1, the expected incremental quantities can be obtained based on the compensation theory, enabling the application of incremental quantity-based supervising elements in IBR-based systems.

To achieve this, two approaches are considered:
a) Modifying the control scheme during the fault to ensure IBR2 is equivalent to IBR1 in the **fault sequence network**, i.e., setting $ne_{ref1} - n^2 i_1 Z_{v1} = ne_{pre1}$. Under this circumstance, based on (8), the magnitude of $Z_{ad}$ becomes 0, and $\Delta Z_{e1}$ is highly inductive. However, the terminal voltage of the IBR during the fault period remains the same as in the pre-fault condition, thereby inherently disabling overcurrent limitation and making this method impractical.
b) Modifying the control scheme during the fault to ensure IBR1 is equivalent to IBR2 in the **pre-fault circuit**, i.e., ensuring $e_{pre1}$ remains equivalent to the IVS ($e_{ref1}$) in series with the virtual impedance ($Z_{v1}$) in the pre-fault circuit.

Fig. 9 (a) illustrates the equivalent source dynamics in the pre-fault circuit. To ensure that IBR1 is equivalent to IBR2 in the pre-fault circuit, $e_{ref1}$ is adjusted such that the terminal voltage of IBR2 ($v_{IBR2}$) matches that of IBR1 ($v_{IBR1}$), i.e. $e_{ref1}-i_{tpre1}Z_{v1}=e_{pre1}$. The IVS during the fault period is adjusted as

$$e_{ref1} = e_{pre1} + i_{tpre1}Z_{v1} \tag{9}$$

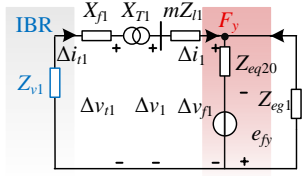

Fig. 10 Pure-fault circuit.

Fig. 9 (b) depicts the equivalent pre-fault circuit after implementing (9) and replacing IBR1 with IBR2. In Fig. 9 (b), the voltage and the current at the relay-assembled point (the bus 1) are the same as those in the pre-fault circuit shown in Fig. 8 (a).

Substituting (9) into (8), the effective impedance for the incremental quantities is derived by

$$\begin{aligned}\Delta Z_{e1} &= jn^2 X_{f1} + jX_{T1} + \frac{ne_{pre1} - (ne_{pre1} + n^2 i_{pre1}Z_{v1}) + n^2 i_1 Z_{v1}}{i_1 - i_{pre1}} \\ &= jn^2 X_{f1} + jX_{T1} + \frac{\overline{ne_{pre1}} - \overline{ne_{pre1}} + n^2 Z_{v1}(\overline{i_1 - i_{pre1}})}{\overline{i_1 - i_{pre1}}} \\ &= jn^2 X_{f1} + jX_{T1} + n^2 Z_{v1}\end{aligned} \tag{10}$$

Based on (10), $\Delta Z_{e1}$ is highly inductive, indicating that the incremental quantity-based supervising elements can operate reliably after adjusting the IVS based on (9).

### B. Pure-Fault Circuit Model

The voltages at the fault location of the fault sequence network, see Fig. 8 (b), and the equivalent pre-fault sequence circuit, see Fig. 9 (b), are separately derived as

$$v_{f1} = \frac{e_{ref1}(Y_{v1}/n^2 + Y_{f1}/n^2 + X_{T1} + mY_{l1}) + v_{g1}Y_{eg1}}{Y_{v1}/n^2 + Y_{f1}/n^2 + X_{T1} + mY_{l1} + Y_{eg1} + Y_{eq20}} \tag{11}$$

$$e_{fy} = \frac{e_{ref1}(Y_{v1}/n^2 + Y_{f1}/n^2 + Y_{f1} + X_{T1} + mY_{l1}) + v_{g1}Y_{eg1}}{Y_{v1}/n^2 + Y_{f1}/n^2 + X_{T1} + mY_{l1} + Y_{eg1}} \tag{12}$$

Subtracting (12) from (11), yields

$$\Delta v_{f1} = v_{f1} - e_{fy}$$
$$= \frac{-e_{y1}Y_{eq20}}{Y_{v1} + Y_{f1} + X_{T1} + mY_{l1} + Y_{eg1} + Y_{eq20}} \tag{13}$$

Fig. 10 presents the pure-fault sequence network based on (13). By adjusting the IVS following (9), the pure-fault circuit of the IBR-based system closely resembles that of SG-based systems shown in Fig. 3 (c), where the load is eliminated.

## V. REALIZATION OF PROTECTION-INTEROPERABLE FRT CONTROL

This section elaborates first on the proposed power control for realizing the idea of the protection-interoperable FRT control. The power control operates in two modes: 1) the protection-interoperable mode, where IVS is adjusted according to (9); and 2) the normal mode, where the IVS is regulated via the normal closed-loop power control. Criteria for switching between the two power control modes are then defined. Finally, a CLC method is introduced to ensure the effective overcurrent limitation in the protection-interoperable mode.

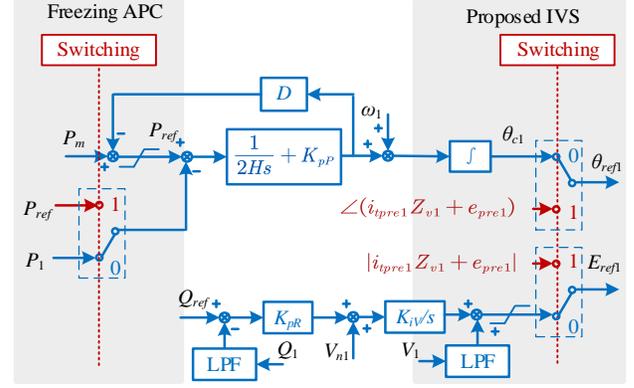

Fig. 11 Proposed power control loop.

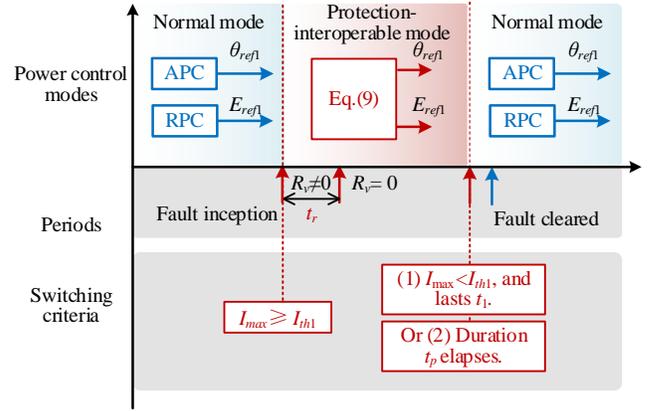

Fig. 12 Control modes for a fault event.

### A. Proposed Power Control

Fig. 11 shows the proposed power control to provide the internal voltage based on (9), where $\omega_1$, $V_{n1}$ are the nominal angular frequency and voltage magnitude, respectively. $D$, $H$, and $K_{pP}$ represent the $P$-$\omega$ droop coefficient, the virtual inertia constant, and the virtual damping constant, respectively. The first-order low-pass filter (LPF) is used in RPC to prevent the impact of measurement noise. Before the fault, the output of the APC ($\theta_{c1}$) is determined by the normal power control, which is expressed as

$$\theta_{c1} = ((P_{ref} - P_1)K_{pP} + \omega_1)\frac{1}{s} \tag{14}$$

After the fault inception, the switches generating $\theta_{ref1}$ and $E_{ref1}$ in Fig. 11 should shift from port 0 to port 1, activating the protection-interoperable mode. In this case, since $P_{ref} \neq P_1$, the resulting error accumulates in $\theta_{c1}$, as described in (14). When the fault is cleared and the switches return to port 0, this accumulated error can cause an undesirable active power transient. To address this issue, the APC is frozen under the protection-interoperable mode by substituting $P_1$ with $P_{ref}$, thereby eliminating the accumulated error in $\theta_{c1}$.

### B. Criteria For Switching Between Power Control Modes

Fig. 12 illustrates the transitions between protection-interoperable and normal power control modes, incorporating the associated switching criteria, where $I_{max}$ is expressed as

$$I_{max} = \max(I_{ta}, I_{tb}, I_{tc}) \tag{15}$$

$I_{ta}$, $I_{tb}$, and $I_{tc}$ represent the peak values of the phase currents for phases $a$, $b$, and $c$, respectively. For a detailed calculation of the peak values, refer to [28]. By comparing $I_{max}$ to a predefined threshold ($I_{th1}$), the system can determine whether to transition into protection-interoperable mode. To enhance sensitivity and ensure reliability of the relay, $I_{th1}$ is set slightly higher than the nominal current $I_{N1}$, e.g., $I_{th1}=1.05I_{N1}$. When $I_{max} \geq I_{th1}$, the protection-interoperable power control is activated.

Once the fault is cleared, the control should transition from protection-interoperable mode back to normal mode. Since the virtual impedance-based CLC method is free of the latch-up phenomenon, the current magnitude can return to the normal level [29]. Consequently, the switching criteria remain based on the POC current magnitude. To prevent frequent switching caused by measurement noises, a time delay of $t_1$ is introduced. If $I_{max}<I_{th1}$ for $t_1$, the system reverts to port 0 in Fig. 11. However, this delay reduces switching sensitivity. To balance reliability and sensitivity, $t_1$ should be set properly. In this paper, $t_1$ is set to 50 ms.

Further, the incremental quantity-based supervising elements store data for a limited time duration, $t_p$ [5]. Once $t_p$ elapses after fault inception, the stored data no longer represent the pre-fault values ($i_{pre1}$ and $v_{pre1}$), causing the magnitude of the incremental quantities to approach zero and deactivating the corresponding supervising elements [11]. Thus, the maximum activation time of the protection-interoperable mode is set to $t_p$, after which the control scheme reverts to normal mode, regardless of the fault clearance status.

### C. Overcurrent Limitation

#### 1) Steady-state overcurrent limitation

Since the zero-sequence current is bypassed by the Δ-Y0 transformer, as shown in Fig. 6, only the positive- and negative-sequence currents are limited to ensure the phase current is within the current limitation, yielding

$$I_{t1} \leq I_{\lim 1}, \ I_{t2} \leq I_{\lim 2} \quad (16)$$

where $I_{lim1}$ and $I_{lim2}$ are the current limitation values.

To limit the positive-sequence current, the corresponding virtual impedance should satisfy

$$\frac{|e_{ref1} - v_{t1}|}{I_{\lim 1}} \leq |Z_{v1} + jX_{f1}| \quad (17)$$

To limit the current and ensure the reliability of supervising elements, substituting (9) into (17) yields

$$\left| \frac{\overbrace{v_{tpre1} + i_{tpre1} jX_{f1}}^{e_{pre1}} + i_{tpre1} Z_{v1} - v_{t1}}{Z_{v1} + jX_{f1}} \right| \leq I_{\lim 1} \quad (18)$$

Based on (18), the conservative requirement for $Z_{v1}$ is derived by

$$|Z_{v1} + jX_{f1}| \geq \frac{|v_{tpre1} - v_{t1}|}{I_{\lim 1} - I_{tpre1}} \quad (19)$$

To limit the negative-sequence current, the requirement for $Z_{v2}$ is derived by

$$\frac{V_{t2}}{I_{\lim 2}} \leq |Z_{v2} + jX_{f2}| \quad (20)$$

Fig. 13 demonstrates the adaptive virtual inductances for positive- and negative-sequence quantities, where $I_{thX1}$ and $I_{thX2}$ denote the current thresholds, beyond which the adaptive virtual inductance is activated. $K_{X1}$ and $K_{X2}$ represent the proportional gains. To mitigate measurement noises, the LPFs, i.e., $G_{LPFI}$, are applied to the measured currents, $I_{t1}$ and $I_{t2}$. To maintain small-signal stability, the first-order LPFs ($G_{LPFX}$) are used with the virtual inductance [19]. The virtual inductances are derived based on the respective current magnitudes, yielding

$$\begin{cases} X_{v1} = K_{X1}(I_{t1} - I_{thX1}) \\ X_{v2} = K_{X2}(I_{t2} - I_{thX2}) \end{cases} \quad (21)$$

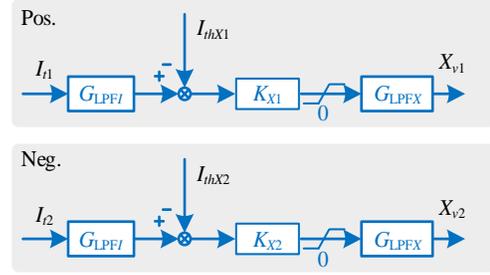

Fig. 13 Adaptive virtual inductance for positive- and negative-sequence quantities.

To ensure the reliability of the supervising elements, only virtual inductances are employed. Combining (19), (20), and (21), the requirements for $K_{X1}$ and $K_{X2}$ are determined as

$$\begin{cases} K_{X1} \geq \dfrac{|v_{tpre1} - v_{t1}|}{(I_{\lim 1} - I_{tpre1})(I_1 - I_{thX1})} - \dfrac{X_{f1}}{I_1 - I_{thX1}} \\ K_{X2} \geq \dfrac{V_{t2}}{I_{\lim 2}(I_2 - I_{thX2})} - \dfrac{X_{f2}}{I_2 - I_{thX2}} \end{cases} \quad (22)$$

Based on (22), the values of $K_{X1}$ and $K_{X2}$ can be set according to the expected maximum values of $|v_{tpre1}-v_{t1}|$ and $V_{t2}$, respectively.

#### 2) Transient overcurrent limitation

To meet the requirements of the supervising elements, the virtual impedance should be highly inductive, which amplifies the DC component decaying rate of the fault current and poses challenges to the transient current limitation [1]. In contrast, virtual resistance accelerates the decay of the DC component in fault currents. Therefore, incorporating virtual resistance can effectively limit transient currents. However, this conflicts with the requirements of protective relays. To address this issue, the transient virtual resistance method in [1] is employed for a maximum duration of $t_r=50$ms, as shown in Fig. 12. At the same time, a time delay ($t_r$) is implemented into the protective relays, which is a common approach to addressing their poor performance during the transient period [30].

## VI. IMPACT ON TRANSIENT STABILITY

This section examines the transient stability of GFM-IBRs with the proposed protection-interoperable control.

First, during the protection-interoperable mode, the APC operates in open loop, eliminating transient stability concerns. However, in the normal mode, the APC transitions to the closed-loop control, affecting the transient stability. To enhance transient stability, the adaptive fast/slow power control method described in [2] can be employed during the normal mode. In this method, the current magnitude is compared with a threshold ($I_{thf}$) to determine whether a fast-timescale power control is needed. When $I_{t1} \geq I_{thf}$, which indicates a huge power angle, the fast-timescale power control is implemented to avoid the loss of synchronization (LOS). Otherwise, a slow-timescale power control is applied to maintain GFM capability. Here, $I_{thf}$=0.94 p.u. according to [2]. For more details on the adaptive fast/slow power control, refer to [2].

Furthermore, the duration of protection-interoperable control mode can be limited by appropriately reducing the value of $t_p$, as shown in Fig. 12. This minimizes the period of continued grid-condition variations under the open-loop power control, thereby mitigating the risk of LOS when the control scheme transitions from the protection-interoperable mode back to the normal mode.

## VII. HARDWARE-IN-THE-LOOP RESULTS

Fig. 14 (a) and (b) depict the actual hardware devices and the topology of the CHIL testing setup, which are used to validate the effectiveness of the proposed FRT method. Three RT-Box3 units are utilized to implement the controller, main circuit, and relay, respectively. Table I presents the parameters of the main circuit. Here, it is assumed an *ag* fault occurs between the bus 1 and the bus 2, *m*=0.01 and the fault resistance is 20Ω. To highlight the effectiveness of the proposed FRT method on the performance of supervising elements, $t_p$ is set to 1s, thereby extending the operating duration of the supervising elements.

Fig. 15 presents the CHIL results of supervising elements under conventional power controls. The correct zones for $\Delta\varphi_1$ and $\Delta\delta_{21}$ are highlighted in blue and grey, respectively, as shown in Fig. 15 (a). When typical power control is used to regulate internal voltage, Eq. (9) does not necessarily hold, leading to deviations of $\Delta\delta_{21}$ and $\Delta\varphi_1$ from their respective bands. Fig. 15 (b) shows the CHIL results for $\delta_{20}$, $\varphi_2$, and $\varphi_0$ where the correct zones are highlighted in grey for $\delta_{20}$ and in blue for $\varphi_2$ and $\varphi_0$. Since the highly inductive virtual impedance-based CLC is implemented and the Δ-Y0 transformer bypasses the direct control of the IBR for zero-sequence quantities, $\delta_{20}$, $\varphi_2$, and $\varphi_0$ consistently fall within their correct bands.

Fig. 16 presents the CHIL results when the proposed FRT method is implemented. In Fig. 16 (a), the correct zones for $\Delta\varphi_1$ and $\Delta\delta_{21}$ are highlighted in blue and grey, respectively. With $E_{ref1}$ and $\theta_{ref1}$ provided based on (9), $\Delta\varphi_1$ and $\Delta\delta_{21}$ remain within the corresponding bands after fault inception, confirming the effectiveness of the proposed FRT method. In Fig. 16 (b), the correct zones for $\varphi_2$ and $\varphi_0$ are highlighted in blue, while the correct zone for $\delta_{20}$ is highlighted in grey. With the implementation of the proposed FRT method, $\varphi_2$, $\varphi_0$, and $\delta_{20}$ remain within their respective bands after fault inception. These CHIL results validate the effectiveness of the proposed protection-interoperable FRT method.

## VIII. EXPERIMENTAL RESULTS

Fig. 17 illustrates the experimental setup used to evaluate the performance of the proposed FRT method during grid disturbances. The 45 kVA Chroma 61850 grid simulator is employed to generate grid disturbances, while the DS1007 dSPACE system is utilized to implement the proposed FRT method. The parameters of the main circuit are provided in Table II.

TABLE I
PARAMETERS OF THE MAIN CIRCUITS FOR CHIL

| Symbol | Meaning | Value |
|---|---|---|
| $v_g$ | Grid voltage (L-L, RMS) | 220kV (1.732 p.u.) |
| $f_1$ | Grid frequency | 50Hz |
| $S$ | Rated power | 100MW (1 p.u.) |
| $n$ | Turn ratio | 33kV/220kV |
| $v_{dc}$ | DC-link voltage | 40kV |
| $L_f$ | Inductance filter | 5 mH |
| $L_T$ | Transformer leakage inductance | 77mH (0.05 p.u.) |
| $l$ | Length of transmission line | 100km |
| $Z_{l1}/l$ | Positive-sequence line impedance | 0.03+j0.34Ω/km |
| $Z_{l0}/l$ | Zero-sequence line impedance | 0.18+j1.19Ω/km |
| $Z_g$ | Grid impedance | 200mH (SCR=6.8) |

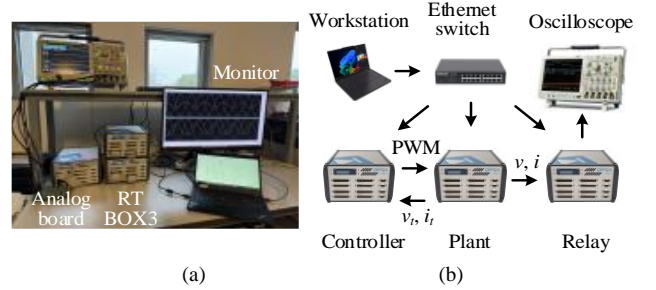

Fig. 14 CHIL testing setup. (a) Hardware devices. (b) Topology.

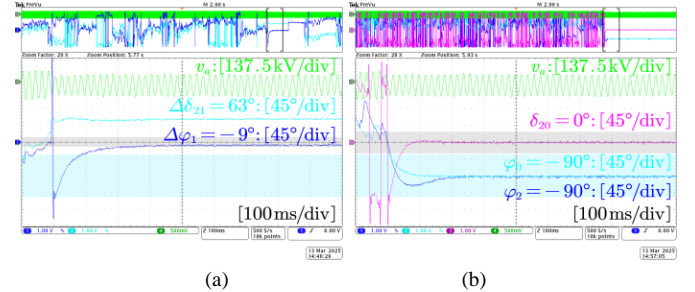

Fig. 15 CHIL results of the supervising elements under conventional power control. (a) Incremental quantity-based elements. (b) Negative-, and zero sequence quantity-based elements.

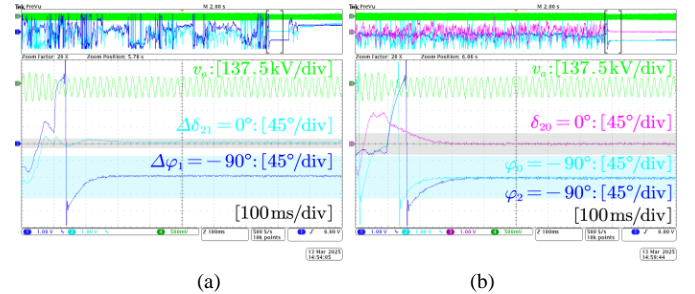

Fig. 16 CHIL results of the protective elements under the proposed FRT method. (a) Incremental quantity-based elements. (b) Negative-, and zero sequence quantity-based elements.

### A. Verification of Switching Criteria

Fig. 18 presents the experimental results validating the switching criteria of the proposed FRT method, where $t_p$=0.3s. Here, if $I_{max} \geq I_{th1}$, $S_f$=1. Otherwise, $S_f$=0. Moreover, when the protection-interoperable mode is activated, $S_p$=1. $\Delta Z_{t1}$ is defined as $\Delta v_{t1}/\Delta i_{t1}$. The red and blue arrows indicate the moments of fault inception and fault clearance, respectively. When $v_g$ drops to 0.7 p.u, the protection-interoperable mode is activated ($S_p$=1), as shown in Fig. 18. During the fault period, $\angle\Delta Z_{t1}$ is close to -90°. After 0.2s, the fault is cleared, and the $S_p$ returns to 0 after $S_I$ remains at 0 for a duration of $t_1$ (50ms). This experimental result validates the effectiveness of the proposed switching criteria.

TABLE II
PARAMETERS OF THE MAIN CIRCUITS FOR EXPERIMENT

| Symbol | Meaning | Value |
|---|---|---|
| $v_g$ | Grid voltage (L-L, RMS) | 86.6V (0.816 p.u.) |
| $f_1$ | Grid frequency | 50Hz |
| $f_s$ | Sampling frequency | 10 kHz |
| $S$ | Rated power | 1kW (1 p.u.) |
| $L_g$ | Grid impedance | 6mH (SCR=4) |
| $L_f$ | Inductance filter | 3mH |

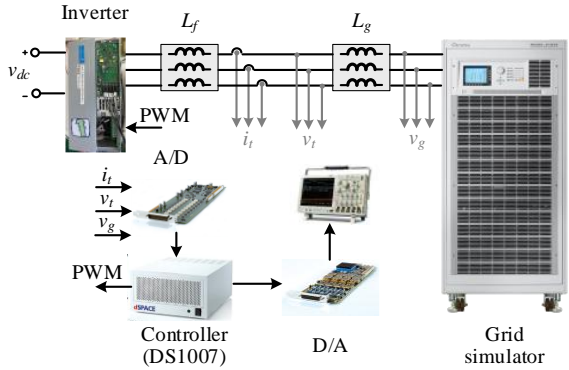

Fig. 17 Diagram of the experimental setup

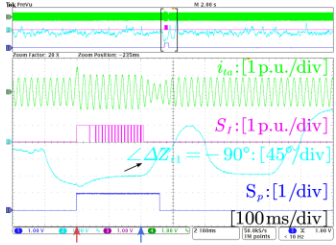

Fig. 18 Effective switching criteria when $v_g$ drops to 0.7 p.u. for 0.2s, and $t_p$=0.3s.

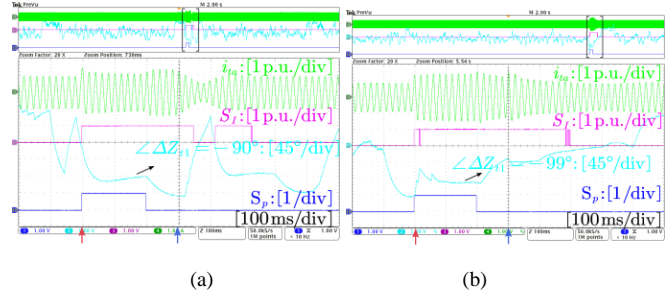

(a)          (b)

Fig. 19 Effective overcurrent limitation under grid disturbances lasting 0.3s, and $t_p$=0.2s. (a) $v_g$ drops to 0.1 p.u. (b) $v_g$ is subjected to a -60° phase jump.

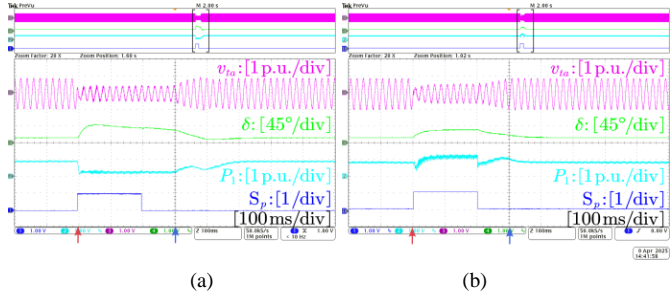

(a)          (b)

Fig. 20 Transient stability maintenance under grid disturbances lasting 0.3s and $t_p$=0.2s. (a) $v_g$ drops to 0.1 p.u. (b) $v_g$ is subjected to a -60° phase jump.

### B. Verification of Overcurrent Limiting Performance

Fig. 19 presents the experimental results for grid disturbances lasting 0.3 s, where $v_g$ drops to 0.1 p.u., as shown in Fig. 19 (a), and a -60° phase jump occurs, as shown in Fig. 19 (b). During the fault period, ∠Δ$Z_{t1}$ remains close to -90° with the implementation of the proposed FRT method. After $t_p$ (0.2s) elapses, the protection-mode control is deactivated, and the control scheme transitions to the normal control mode. Under severe voltage sag and phase jump conditions, the fault current remains below 1.5 p.u. These experimental results confirm the effectiveness of the proposed FRT method in limiting fault current during severe grid disturbances.

### C. Verification of Transient Stability

Fig. 20 presents the experimental results evaluating the impact of the proposed FRT method on transient stability, with $t_p$=0.2s. Here, $δ$ is the angle difference between the POC and grid voltages. In Fig. 20 (a), $v_g$ drops to 0.1 p.u. for a duration of 0.3s. In Fig. 20 (b), a -60° phase jump occurs and lasts for 0.3s. These experimental results show that under the implementation of the proposed FRT method, the IBR maintains transient stability under severe grid disturbances.

## IX. CONCLUSION

This paper proposes a protection-interoperable FRT method to enhance the performance of the incremental quantity-based supervising elements in IBR-based systems. By modifying the power control and CLC of the GFM-IBR, the proposed method not only improves the reliability of incremental quantity-based supervising elements during the short circuit faults but also ensures effective overcurrent limitation and transient stability under grid disturbances. The effectiveness of the proposed method is validated through CHIL tests for supervising elements. Furthermore, its performance under severe grid disturbances is confirmed through experimental tests.